\documentclass[sigconf]{acmart}
\usepackage[T1]{fontenc}
\usepackage[utf8]{inputenc}
\usepackage{proof}
\usepackage{textcomp}
\usepackage{amsmath}
\usepackage{amsthm}
\usepackage{amssymb} 
\usepackage{xparse} 
\usepackage{xargs}
\usepackage{xspace}
\usepackage{hyperref}
\usepackage{mathtools}
\usepackage{graphicx}
\usepackage{algorithm,algorithmicx,algpseudocode}
\makeatletter 
\algnewcommand{\LineComment}[1]{\Statex \hskip\ALG@thistlm \(\triangleright\) #1}

\newcommand{\paran}[1]{\left( #1 \right)}

\title{A Most Irrational Foraging Algorithm}

\author{Abhinav Aggarwal}
\affiliation{Department of Computer Science\\University of New Mexico Albuquerque, USA}
\email{abhiag@unm.edu}

\author{William F. Vining}
\affiliation{Moses Biological Computation Lab\\University of New Mexico Albuquerque, USA}
\email{wfvining@cs.unm.edu}

\author{Diksha Gupta}
\affiliation{Department of Computer Science\\University of New Mexico Albuquerque, USA}
\email{dgupta@unm.edu}

\author{Jared Saia}
\affiliation{Department of Computer Science\\University of New Mexico Albuquerque, USA}
\email{saia@cs.unm.edu}

\author{Melanie E. Moses}
\affiliation{Moses Biological Computation Lab\\University of New Mexico Albuquerque, USA}
\email{melaniem@unm.edu}

\newcommand{\goldenfa}{\textsc{GoldenFA}\xspace}

\setcopyright{acmcopyright} 
\acmDOI{...} 
\acmISBN{...} 
\acmYear{2019} 
\copyrightyear{2019} 
\acmConference[BDA'19]{The 7th Workshop on Biological Distributed Algorithms (BDA 2019)}{July 29, 2019}{Toronto, CA.}



\begin{document}
\begin{abstract}
We present a foraging algorithm, \goldenfa, in which search direction is chosen based on the Golden Ratio. We show both theoretically and empirically that \goldenfa is more efficient for a single searcher than a comparable algorithm where search direction is chosen uniformly at random. Moreover, we give a variant of our algorithm that parallelizes linearly with the number of searchers.  
\end{abstract}

\maketitle

\section{Introduction}
\emph{"Golden Ratio is a powerful mathematical constant woven into the very fabric of biology. It is the unique visual tension between comforting symmetry and compelling asymmetry, and its thoughtful application can bring beauty and harmony and intrigue to all manner of designed things."} - Darrin Crescenzi\\

The Fibonacci sequence ${1,1,2,3,5,8,13,...}$ is created by the recurrence $F_n = F_{n-1} + F_{n-2}$ and is one of the most famous biologically-inspired mathematical sequences. The Golden Ratio, denoted $\phi$, is the limit of the ratio of consecutive numbers in this sequence. Fibonacci generated the sequence as an idealized model of a reproducing rabbit population assuming overlapping generations~\cite{dunlap1997golden}. It was documented in India many centuries earlier, and has been observed in numerous biological systems including the arrangement of pine cones, unfurling of fern leaves, and the arrangement of sunflower seeds that optimally fills the circular area of the flower~\cite{naylor2002golden}. The Golden ratio and Fibonacci numbers have been used in computer science for various applications like obtaining optimal schedules for security games~\cite{kempe2018quasi}, Fibonacci hashing~\cite{knuth1973art}, bandwidth sharing~\cite{itai1984golden}, data structures~\cite{fredman1987fibonacci} and game theoretic models for blocking-resistant communication~\cite{king2011conflict}.

In this paper, we use the Golden Angle created by arcs that form the Golden Ratio to develop a collective foraging algorithm that reduces the time to first discovery of clustered targets. In particular, we show that our new foraging algorithm, \goldenfa, performs better in theory and practice than a previous algorithm that chooses search directions uniformly at random~\cite{aggarwalIROS2019}.

\vspace{0.3em}\noindent \textbf{Motivation for using the ``most irrational" number.}  Any number can be written as an integer plus $1$ over another number. The larger the denominator in the fractional part, the better the integer part is as an approximation.  For example,  $\pi = 3 + 1/x$, for $x \geq 7$, and so $\pi$ is fairly-well approximated by $3$.  Thinking recursively, the number $x$ in the denominator can itself be written as an integer plus $1$ over another number.  Thus, we can write any number as a (possibly infinite) continued fraction~\cite{khinchin1963continued} $$x_1 + \frac{1}{x_2 + \frac{1}{x_3 + \ldots}}$$ where the $x_i$ values are all integers for $i \geq 1$.  

The degree to which the original number is well-approximated by a finite continued fraction depends on how large the $x_i$ values are.  When  $x_i = 1$ for all $i \geq 1$, we obtain an irrational number that is most difficult to approximate.  To find this most difficult irrational number, we set $y = 1 + \frac{1}{y}$, and solve the resulting quadratic equation to obtain a solution $y = \frac{1 + \sqrt{5}}{2}$, which is the celebrated Golden ratio $\phi$.

The fact that $\phi$ is difficult to approximate with a rational number has useful implications in ensuring angles are ``well-spread".  For example, if we start at the point $0$ on a unit circle, and iteratively add points by moving clockwise by distance $\phi$, then we will end up with points that spread out with ample distance between points.\footnote{In contrast, if we use $\pi$ instead of $\phi$, the points will cluster around $3$ spokes since $\pi$ is fairly well-approximated by $3$.  See~\cite{numberphilevideo} for a fascinating simulation and discussion of these facts.}  Interestingly, this approximates how plants add florets, leaves and petals as they grow. If the next leaf is added by moving distance $\phi$ along a unit circle, this ensures that leaves are well-spread in order to increase their total intersection with sunlight without interference. 

In this paper, we make use of the ``most irrational" property of $\phi$ to design a foraging algorithm.  A searcher first forages in an initial direction from a nest site to a boundary of the search space, and then returns to the nest site.  The angle for the next spoke from the nest site is chosen by moving a distance of $\phi$ along a unit circle.  In this way, we can ensure that our foraging spokes are ``well-spread", thus minimizing overlap at the circle center while maximizing the probability of the first discovery of a cluster of resources somewhere in the circle area.  If there are multiple searchers, it is straight-forward to parallelize this process.

The rest of the paper is organized as follows:  We first define our formal model and problem statement.  Then we introduce \goldenfa and explain the theoretical upper bound on finding a single target of a given diameter for a single searcher and then multiple searchers. We then compare theoretical predictions to simulated searchers.

\section{Our Model and Problem Statement}
We assume a circular arena of radius $R$ with the \emph{nest} at the centre\footnote{Similar to existing work, this arena can be modelled as a discrete grid with the Manhattan distance as the metric, however, since working in the continuous Euclidean space introduces only a constant multiplicative blowup, we state our algorithm and the results for a continuous arena for mathematical simplicity.}. A \emph{cluster}\footnote{We assume that this cluster is circular in shape for mathematical simplicity, however, our results apply to all cluster shapes that can be circumscribed by a circle.} of targets is placed at a distance $D$ from this collection zone and has a diameter $\Delta$.
We assume an obstacle-free arena in which the cluster does not move or regenerate as targets are collected.

We assume $N \geq 1$ \emph{searchers}. Each searcher has limited memory and can complete straight line motion in a specified direction from the nest to the edge of the search arena. Searchers can detect if they have encountered a target. Moreover, they know the direction in which they are currently moving, but no information from the past can be stored and or communicated. 

\vspace{0.3em}\noindent \textbf{Overview of our Approach.} 
Previous work shows that foraging for a single target in an arena without any knowledge of the arena parameters requires time that is proportional to the area of the arena~\cite{feinerman2017ants}. We seek to reduce time to discover a  cluster with large diameter. In particular, when cluster diameter is any increasing function of arena diameter our foraging strategy can locate the cluster in sub-quadratic time.  

We state our main result in the theorem below and defer its proof to the full version of our paper.

\begin{theorem}
\label{thm:mainGoldenFA}
The number of time steps taken by \goldenfa before the cluster is located for the first time is $O\paran{\paran{\frac{R}{N\Delta}+1}D}$.
\end{theorem}
\noindent  Again, as long as $\Delta$ is some increasing function of $D$, \goldenfa has improved performance.  Additionally, when $N = R$ searchers work in parallel, the distance travelled is $O(D)$, which is asymptotically optimal. 




Technically, the proof of this theorem rests on two main facts.  First, $\sin^{-1}x = \Theta(x)$ for $x<1$ by the Taylor expansion, so the angle formed with the nest and cluster diameter is of size $\Theta(\Delta/D)$.  Second, to ensure the largest angle size between consecutive spokes is less than $\Theta(\Delta/D)$, our algorithm requires adding $O(D/\Delta)$ spokes by the Three-Gap Theorem~\cite{swierczkowski1958successive}.  Thus, no matter where the cluster is located in the arena, we require $O(D/\Delta)$ spokes before some spoke intersects the cluster. 

\section{The Golden Foraging Algorithm}
We now describe our algorithm, \goldenfa.

\vspace{0.3em}\noindent \textbf{Single-searcher case:}  
The searcher starts from the reference heading, moves along a  spoke to the end of the arena and then returns to the nest. Next, the searcher turns an arc distance of $\phi$ along a unit circle centered at the nest, and moves along a spoke in that direction to the end of the arena.  This process continues until the searcher locates the cluster.

\begin{figure}[t]
    \centering
    \includegraphics[width=\columnwidth, trim={15cm 6.8cm 2cm 6.8cm},clip=true]{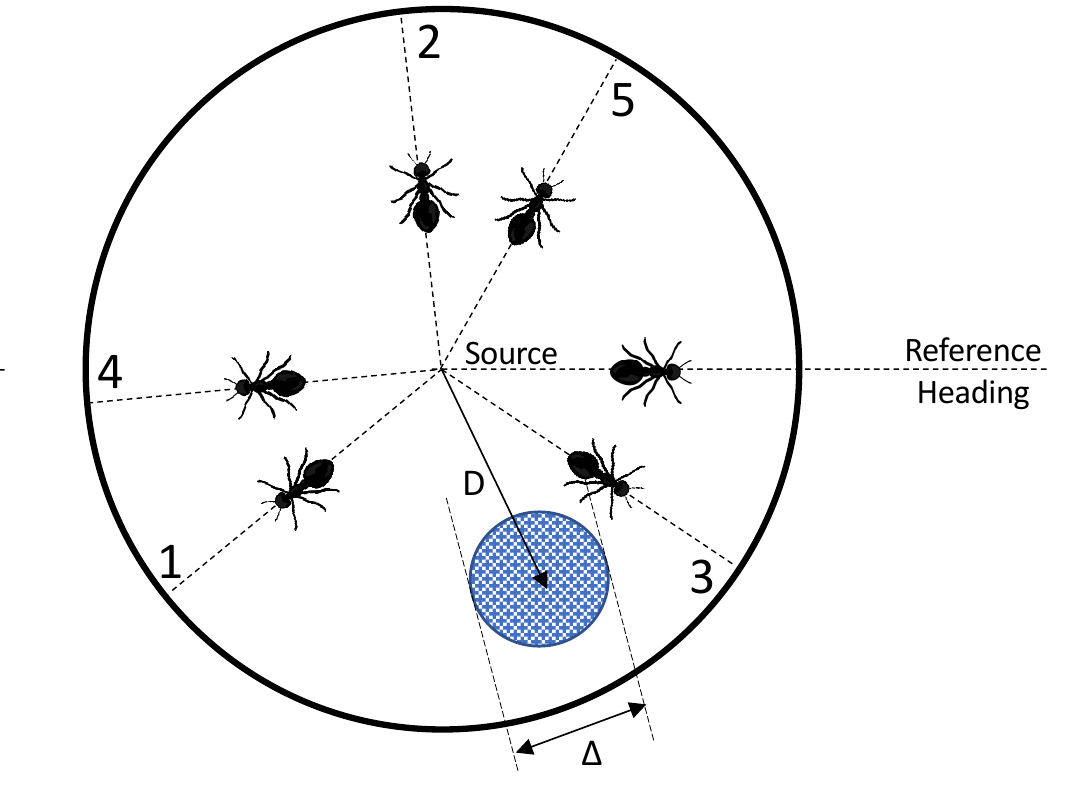}
    \caption{A schematic of the \goldenfa algorithm for a single cluster (shown as the shaded area) of diameter $\Delta$ located at a distance $D$ from the source.}
    \label{fig:goldenFA_schematic}
\end{figure}

\vspace{0.3em}\noindent \textbf{Multiple searchers case:}
When $N\geq 1$ searchers search for the cluster together, searcher $i$ starts the search at an arc distance of $(i-1) \phi$ from the reference heading and searches along spokes that are $N \phi$ arc distance apart (see Figure~\ref{fig:goldenFA_schematic}). 

For this to work, each searcher must have a unique identifier and know its relative order in the sorted sequence of these identifiers. For simplicity, we assume that the search stops as soon as any searcher discovers a target. 

\section{Experimental Evaluation}
The theoretical analysis in the previous section makes many assumptions that do not hold in a robotics setting. These include the lack of collisions between robots, which is a cause of significant slow-down in central place  foraging~\cite{Lu2016mpfa}; the use of an exact value of $\phi$ instead of an approximation; and imprecise robot motion. To test whether the analysis holds under more realistic constraints, we implemented the multiple-searcher \goldenfa in the ARGoS swarm robotics  simulator~\cite{pinciroli2012argos}, and evaluated the time required to discover a single square cluster of resources placed uniformly at random in a $100 \times 100$ meter square arena. 

The resources in the cluster are placed 0.15 meters apart such that a square cluster with $k$ resources on a side has diameter  $\Delta \approx 0.15k$ meters. We conducted experiments with square clusters of $8\times 8$, $16\times16$, $32\times 32$, $64\times 64$, and $128\times128$ resources. For each cluster and value of $N$, we computed the average time over 100 experiments.

\begin{figure}
    \centering
    \includegraphics[width=\linewidth]{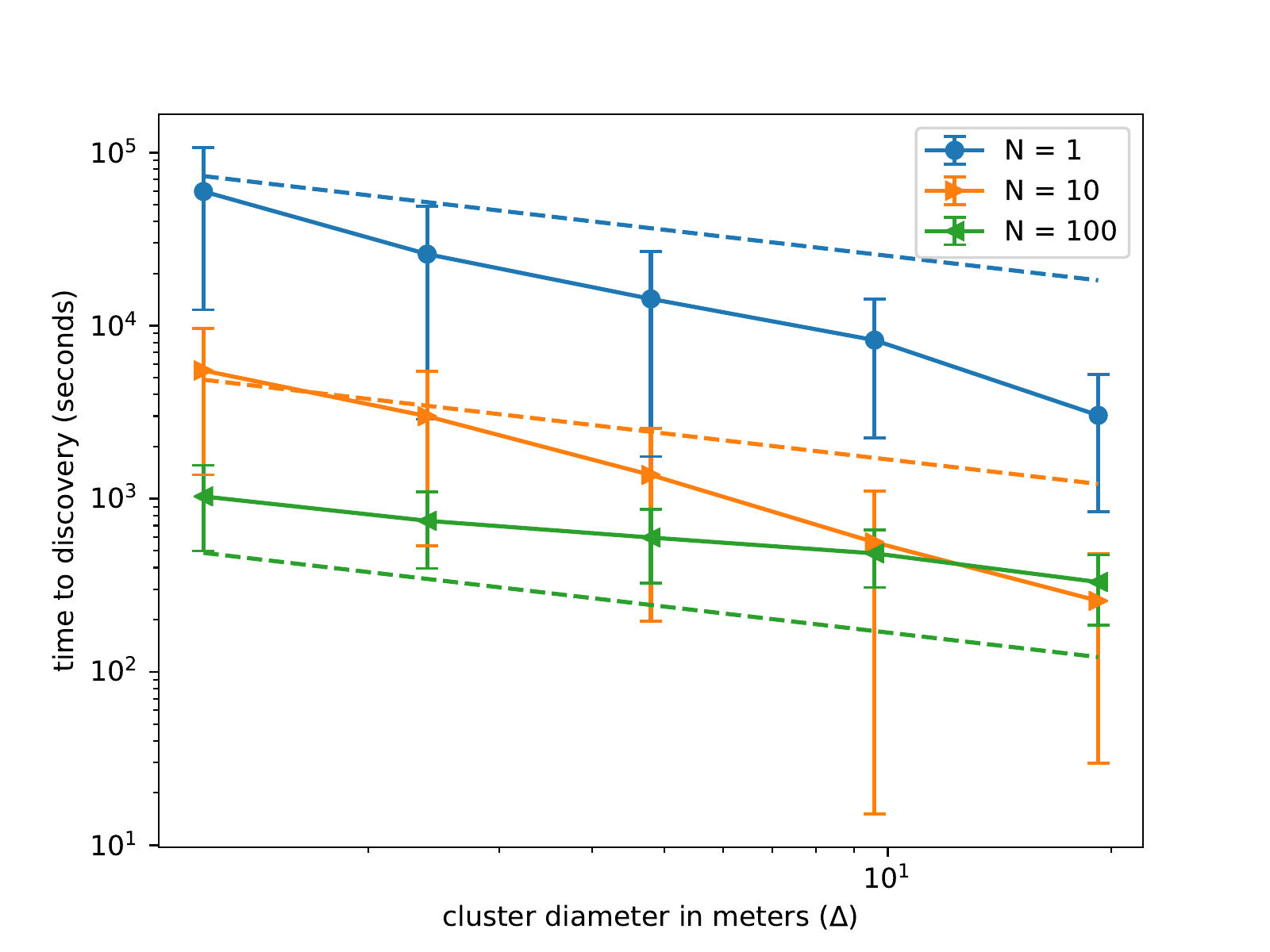}
    \caption{Mean time to discover the single cluster in a $100 \times 100$ meter arena. The dashed lines show the predicted time to discover the cluster $f(x) = 32(\frac{RD}{N\Delta})$ (from Theorem~\ref{thm:mainGoldenFA}) where $D = \frac{2}{3}R$ (the expected distance of a randomly placed cluster from the nest). Error bars show plus or minus one standard deviation. As the swarm size increases clusters are found more quickly; however, for $N=100$ the effects of congestion near the nest slow down the search and lead to longer than the predicted search time.}
    \label{fig:golden-fa-performance}
\end{figure}

\vspace{0.3em} \noindent \textbf{Time to Discover versus $\Delta$ and N.}
Figure~\ref{fig:golden-fa-performance} shows how the time to discover decreases as $\Delta$ increases.  There seems to be a nice linear decrease, as predicted in Theorem ~\ref{thm:mainGoldenFA}, for $N=1$ and $N=10$, with a slope of $32$.  However, for $N=100$, the slope is larger, probably because of congestion effects.  We note that congestion happens when searchers are not infinitely small, as is assumed by the theory. 

\begin{figure}
    \centering
    \includegraphics[width=\linewidth]{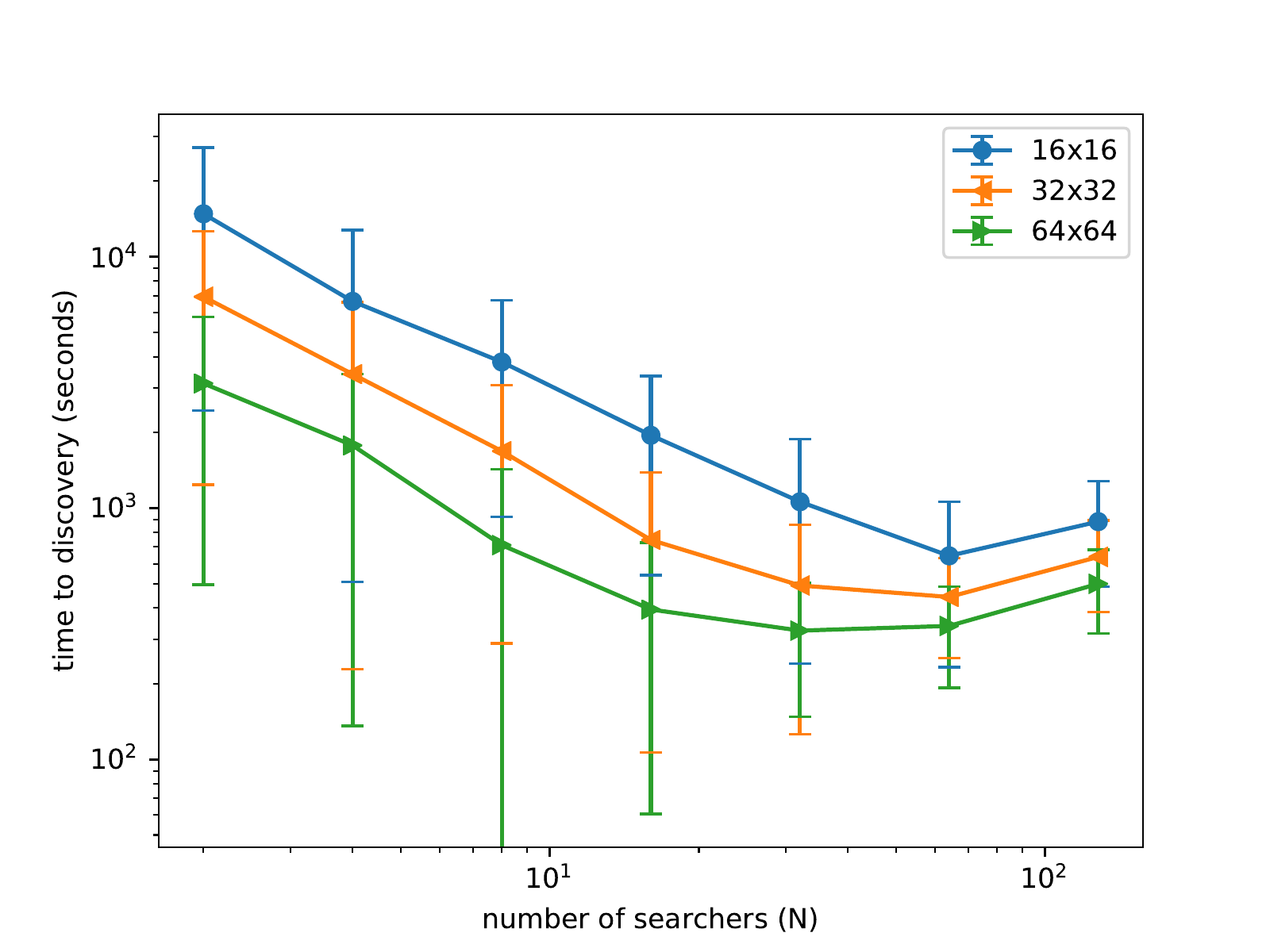}
    \caption{Mean time to discover the cluster as the number of searchers increases for three cluster diameters. The effect of increased congestion is clearly visible as an increase in time to discovery in the largest swarms.}
    \label{fig:scale-swarm}
\end{figure}

Figure~\ref{fig:scale-swarm} shows how the time to find the cluster depends on the number of searchers.  For increasing $N$, we evaluated square clusters of size $16\times16$, $32\times 32$ and $64\times 64$  in a $100 \times 100$ meter arena. The figure shows how additional searchers reduce the search time, but how adding too many searchers causes increased congestion, which results in increased search times.

\vspace{0.3em} \noindent \textbf{Comparison to Ballistic Algorithm.} We compared our algorithm with the Ballistic Central Place Foraging Algorithm (b\textsc{CPFA})~\cite{aggarwalIROS2019}, for the case $N=10$. In b\textsc{CPFA}, the searchers traverse spokes in directions  chosen uniformly at random, until the cluster is located. It can be shown that the number of spokes required by this algorithm is $O\paran{\frac{D}{\Delta}\log\frac{D}{\Delta}}$ with high probability, using the result about maximum distance between random points on an arc from~\cite{king2004choosing}.  This is asymptotically worse than \goldenfa for all values of $D$ and $\Delta$.

\begin{figure}
    \centering
    \includegraphics[width=\linewidth]{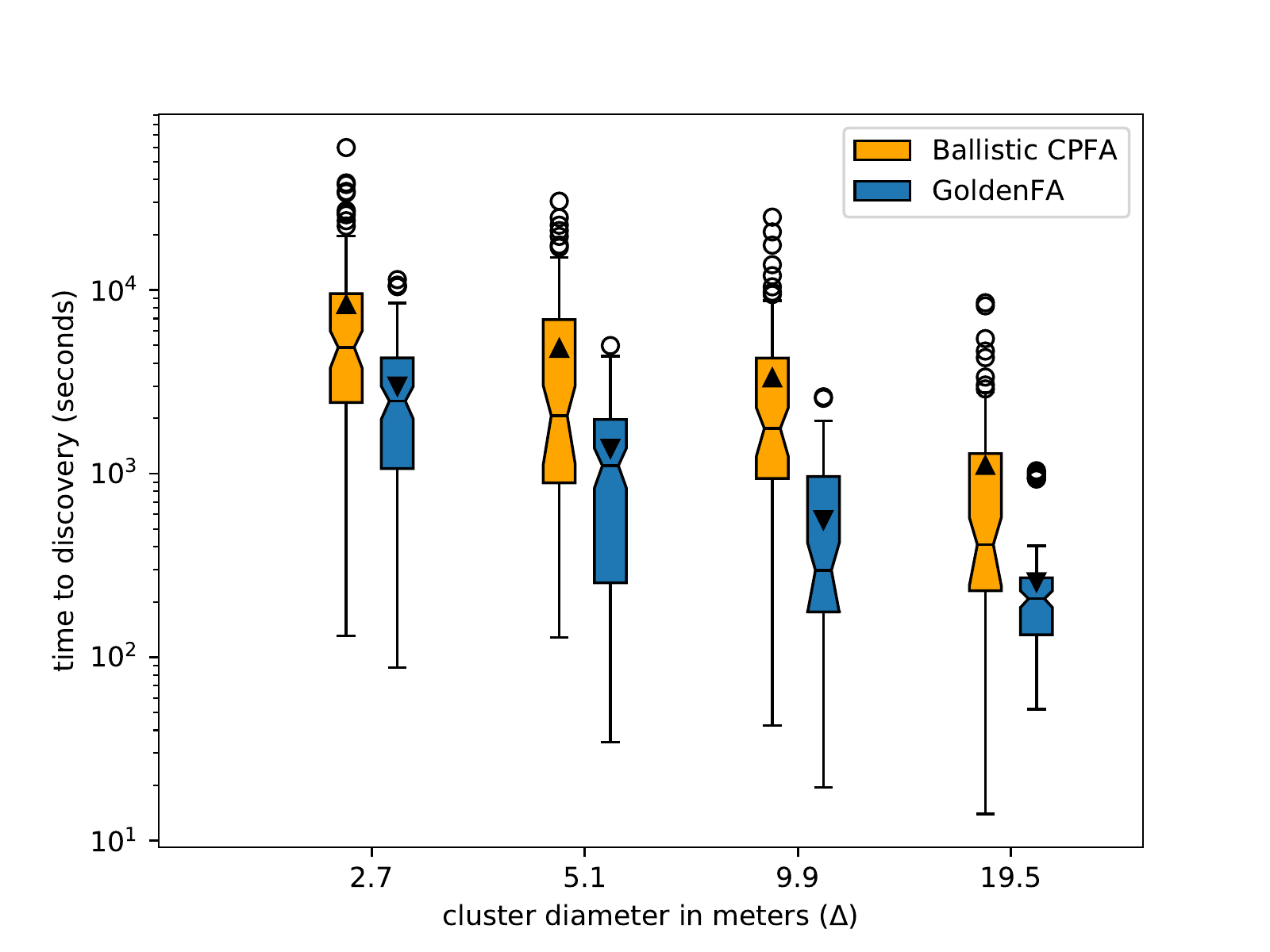}
    \caption{Comparison of b\textsc{CPFA} and \goldenfa. The time for 10 searchers to find a single cluster in a $100\times100$ meter arena as the cluster diameter increases. Black triangles indicate the mean. Not only does it take longer to find the cluster using the ballistic \textsc{CPFA}, but the search time is more variable compared to \goldenfa.}
    \label{fig:bcpfa-vs-golden}
\end{figure}

Figure~\ref{fig:bcpfa-vs-golden} shows that the searchers using \goldenfa are able to find the cluster faster than those using the Ballistic \textsc{CPFA} (b\textsc{CPFA}). The results for varying cluster diameter show that the search time using b\textsc{CPFA} is both longer and more variable than the search time using \goldenfa. Ballistic \textsc{CPFA} searches have a large number of outliers that take considerably longer than average time to find the cluster.

\section{Related Work}
Search is a fundamental problem in  biology, where survival depends on search for mates, prey and other resources.  It is also a common problem in robotics and mobile computing.  Collective search, where multiple searchers must coordinate, is a key problem in computer science, robotics and in social insects. Ant- and bee-inspired algorithms have been particularly influential in swarm robotics research~\cite{csahin2004swarm, krieger2000ant, karaboga2009survey}.  In prior work, we have used algorithms inspired by foraging behaviours of desert seed-harvesting ants. These ants forage collectively as follows : each ant leaves a central nest, travels in a relatively straight line in an apparently randomly chosen direction.
Upon finding food, the ant determines whether to remember and return to that location or communicate the location to nest mates by laying a pheromone trail.

We mimic this behaviour in robots with a generic Central Place Foraging Algorithm (CPFA) which is effective at finding nearby resources quickly, particularly when resources are distributed in multiple clusters of unknown diameter~\cite{Hecker2015}. However, a simpler interlocking spiral algorithm finds targets faster than the bio-inspired CFPA, and the spiral is particularly fast at completely collecting all targets~\cite{heckerClusters, Fricke2016b, aggarwalIROS2019}. 


Our prior work~\cite{aggarwalIROS2019} shows that the most efficient foraging algorithms that completely retrieve all items in a search arena in minimal time use geometric patterns that fill space with minimal traversal of already searched territory. In contrast, in this work, we seek foraging patterns that minimize the time to first discovery when resources are clustered in a single pile of unknown diameter.

\section{Discussion}
\vspace{0.3em} \noindent \textbf{Foraging for Multiple Clusters} When multiple clusters are placed in the arena, time for first detection will be determined by the largest cluster. For discover of all clusters, the number of spokes must be large enough to find the cluster with smallest diameter.  Note that if the smallest cluster has a diameter that is constant, then the asymptotic performance of \goldenfa is no better than exhaustive search, which has quadratic cost. 

\vspace{0.3em} \noindent \textbf{Conclusions and Future Work:} We have described an algorithm, \goldenfa that can locate a cluster efficiently.  Our algorithm has search time that asymptotically decreases linearly with cluster diameter and the number of searchers.  Moreover, our algorithm performs well in practice, with search times that generally match our theoretical predictions.  Further, our algorithm empirically outperforms another common search algorithm in both mean time to discovery and variance.  We believe our algorithm is a first step toward minimizing foraging time when resources are distributed in an unknown number of clusters of unknown and variable diameter.

 Next steps of this work include the following.  First, implementing \goldenfa in real, rather than simulated, robots to determine how well it performs with real-world noise and error. Second, a theoretical analysis of resilience of \goldenfa. Finally, measuring collection time of multiple clusters by \goldenfa and comparing to other foraging algorithms. 


\bibliography{ref}
\bibliographystyle{acm}
\end{document}